# Solving multiple-criteria R&D project selection problems with a data-driven evidential reasoning rule

Fang Liu
School of Accounting, Zhejiang University of Finance and Economics, Hangzhou, 310018, Zhejiang, China

Yu-wang Chen
Alliance Manchester Business School, The University of Manchester, Manchester M15 6PB, United Kingdom

Jian-bo Yang
Alliance Manchester Business School, The University of Manchester, Manchester M15 6PB, United Kingdom

Dong-ling Xu
Alliance Manchester Business School, The University of Manchester, Manchester M15 6PB, United Kingdom

Wei-shu Liu
Corresponding author
wsliu08@163.com
School of Information Management and Engineering, Zhejiang University of Finance and Economics, Hangzhou 310018, Zhejiang, China

**Abstract:**
   In this paper, a likelihood based evidence acquisition approach is proposed to acquire evidence from experts' assessments as recorded in historical datasets. Then a data-driven evidential reasoning rule based model is introduced to R&D project selection process by combining multiple pieces of evidence with different weights and reliabilities. As a result, the total belief degrees and the overall performance can be generated for ranking and selecting projects. Finally, a case study on the R&D project selection for the National Science Foundation of China is conducted to show the effectiveness of the proposed model.
   The data-driven evidential reasoning rule based model for project evaluation and selection (1) utilizes experimental data to represent experts' assessments by using belief distributions over the set of final funding outcomes, and through this historic statistics it helps experts and applicants to understand the funding probability to a given assessment grade, (2) implies the mapping relationships between the evaluation grades and the final funding outcomes by using historical data, and (3) provides a way to make fair decisions by taking experts' reliabilities into account. In the data-driven evidential reasoning rule based model, experts play different roles in accordance with their reliabilities which are determined by their previous review track records, and the selection process is made interpretable and fairer. The newly proposed model reduces the time-consuming panel review work for both managers and experts, and significantly improves the efficiency and quality of project selection process. Although the model is demonstrated for project selection in the NSFC, it can be generalized to other funding agencies or industries.

**1. Introduction**

With the rapid development of science and technology, the problem of selecting research and development (R&D) projects is becoming increasingly important (Arratia M et al., 2016; Pinheiro et al., 2016; Santamaría et al., 2010). Many countries have established specific research funding agencies and designed formal procedures to evaluate and select projects. R&D project evaluation and selection is a frequent process and a significant task for funding agencies (Karasakal and Aker, 2016; Silva et al., 2014; Tang et al., 2017). As a typical multiple-criteria decision analysis problem, qualitative information often needs to be used in the selection process. The assessments from different experts can have different weights and reliabilities, which can significantly influence the decision analysis for project selection. Due to the exploding number of alternative proposals, the natures of R&D projects, and the subjective judgements of experts involved in the selection process, it has become especially critical and challenging for agencies to make rational and informative project funding decisions (Tavana et al., 2013).

R&D project evaluation and selection is a complicated multi-criteria decision-making process (Chiang and Che, 2010; Eilat et al., 2008). The decision makers have to determine which new proposals should be chosen for implementation. To make the decision making process transparent and consistent, research funding agencies tend to follow a structured, formalized decision process and select projects in a consistent way (Huang et al., 2008).

R&D project selection process can be carried out through several steps, namely, call for proposals, proposal submission, proposal validation, peer review, assessments aggregation, panel review and final decision (Feng et al., 2011; Silva et al., 2014; Zhu et al., 2015). Studies by Silva et al. (2014) revealed that the current approaches for R&D project selection either automate workflows or analyze only activities such as proposal clustering, reviewer assignment, and portfolio evaluation. Rather limited work is focused on aggregation methods for R&D project selection (Liu et al., 2017). The aggregation of assessment information serves as one of the key steps in the R&D project selection process. At this step, managers disseminate a comprehensive evaluation result for a project according to the rules and policies, as well as the assessments of the project provided by experts. This step includes the aggregation of multiple experts' assessments on multiple evaluation criteria, which is critical for the final decision. There are mainly three elements that need to be taken into account when aggregating assessments from experts: representing assessments, measuring their weights and reliabilities, and dealing with conflicting assessments. The three elements should be handled differently in the light of the circumstances.

In order to deal with assessments aggregation issues and underpin the above three elements, the paper presents a data-driven inference model based on the evidential reasoning (ER) rule for supporting group-based project selection decisions (Yang and Xu, 2013; 2014). The ER rule constitutes a general conjunctive probabilistic inference process and has the features of handling highly or completely conflicting evidence rationally. The proposed data-driven inference model includes the following two main components: likelihood based evidence acquisition to handle experts' subjective judgments (qualitative information) and transform them into multiple pieces of evidence, and the aggregation of information by multiple experts and multiple criteria by using the ER rule.

The rest of this paper is organized as follows. In Section 2, we review the existing literature. In Section 3, the ER rule is briefly described and the details of the proposed method are presented. A case study for R&D project selection for the National Science Foundation of China (NSFC) is described and the evaluation results are analyzed and presented in the fourth section. It is followed by the concluding remarks section where key research findings and further work are summarized and discussed respectively.

## 2. Literature review

### 2.1 Challenges of peer review

Peer review is widely seen as vital to R&D project selection and has a broad application in practice for many years. Despite the importance of peer review, it has been subject to intense criticism for various kinds of bias and for rendering unfair outcomes. Relevant studies of peer review have substantiated that assessments can be influenced by various kinds of bias, including scholarly bias and cognitive bias (Langfeldt, 2006). According to Chubin and Hackett (1990), the outcomes of peer review can be influenced by cronyism and scientific feuds. Institutional particularism can lead to unjust outcomes of peer review, for example, when the reviewers come from a similar or similar type of institution with a proposal to be judged, they tend to support this proposal because of the institutional similarity (Travis and Collins, 1991; Luukkonen, 2012). Furthermore, assessments could be inflated because of optimism bias, as some experts may be too lenient in their assessments for various reasons (Wang et al., 2005).

Besides bias, assessments can also be influenced by other factors. For instance, limited cognitive horizons of experts and inherent nature of various R&D projects make the selection process even more difficult as no single expert could understand all of the submitted proposals (Wang et al., 2005). The abilities to provide correct assessments by multiple experts can be very different. Since experts have different abilities to provide valuable assessments, the funding results of synthesizing assessments could be deviated from actual intentions of experts if their assessments are treated equally.

Instead of avoiding such questions through external or institutional constraints, one more overall question can be addressed when using peer review: reliability measuring the quality of assessments can be taken into account during selection process. It is important to identify reviewers who produce reliable reviews because reliable reviews should improve funding decisions about which proposals to fund and should help improve projects that are eventually funded. From this perspective, to improve their reliabilities, experts will have to keep objective and try to make high quality recommendations, so as to make the evaluation process reliable and fair.

**2.2 Studies on R&D project selection methods**

In the past decades, various analytical methods and techniques have been developed to support better decision making in R&D project selection, ranging from qualitative review to quantitative mathematical programming. Comparative studies on the methods of R&D project evaluation and selection have been conducted by Baker and Freeland (1975), Jackson (1983), Henriksen and Traynor (1999), Poh, Ang and Bai (2001), Meade and Presley (2002). Moreover, there are several different R&D project evaluation methods adopted by researchers: mathematical programming (Badri et al., 2001), decision support system (Tian et al., 2005), fuzzy multiple attribute decision-making method (Wei et al., 2007), analytic network process (Jung and Seo, 2010), data envelopment analysis (Ghapanchi et al., 2012; Karasakal and Aker, 2016) and artificial neural networks (Costantino et al., 2015). Feng et al. (2011) and Tavana et al. (2015) use integrated methods that integrate multiple methods such as AHP, scoring and data envelopment analysis to support R&D project evaluation decisions.

The usual approaches suffer from a number of deficiencies, ranging from problems in methodology (such as treatment of uncertainty) to more fundamental concerns with the overall approach (Liu et al., 2017). For example, AHP suffers from the deficiency of 'rank reversal problem' and experts may also face the serious challenge of large number of pairwise comparisons (Poh et al., 2001). In addition, most studies focus on describing the mechanisms of the techniques, or analyzing their strengths and weaknesses based on the nature of R&D project (Hsu et al., 2003; Poh et al., 2001). The afore-mentioned methods cannot deal with uncertainties from subjective judgment and may ignore the organizational decision process (Schmidt and Freeland, 1992). Adopting such idealized vision based models requires organizations to change the process in which they currently make project selection decisions. Few have gained wide acceptance for real-world R&D project selection (Tian et al., 2005).

**2.3 R&D project selection using the evidential reasoning rule**

The evidential reasoning (ER) rule is developed for conjunctively combining multiple pieces of independent evidence with weights and reliabilities, and it is a new development of seminal Dempster-Shafer (D-S) evidence theory (Shafer, 1976), the ER algorithm (Xu, 2012; Yang and Singh, 1994; Yang and Xu, 2002) and decision theory. The ER rule takes into account both the bounded sum of individual support and the orthogonal sum of collective support for a hypothesis when combining two pieces of independent evidence, and it constitutes a generic conjunctive probabilistic reasoning process (Yang and Xu, 2013; 2014). Based the orthogonal sum operation, the ER rule is inherently associative and commutative and can be used to aggregate multiple pieces of evidence in any order.

In the ER rule, a frame of discernment is defined by a set of hypotheses that are mutually exclusive and collectively exhaustive. A piece of evidence can be profiled by a belief distribution (BD) on the power set of the frame of discernment, where basic probabilities can be assigned to not only singleton hypothesis but also any of their subsets. BD is the natural generalization of conventional probability distribution in which basic probabilities are assigned to singleton hypotheses only. It has been shown that the ER rule is equivalent to Bayesian' rule if likelihoods generated from sample data are normalized to acquire basic probabilities in the ER framework (Yang and Xu, 2014). By equivalence, it means that Bayesian inference can be precisely conducted yet in a symmetrical process in the ER paradigm where each piece of evidence is profiled in the same format of probability distribution. Zhu et al. (2015) and Liu et al. (2017) uses the belief distributions to represent evaluation information and employs the ER rule to combine multiple pieces of evaluation evidence.

It appears that the ER rule is an ideal tool for multi-criteria performance evaluation with taking into account weight and reliability. R&D project selection for research funding agencies requires the consistency of evaluation standard and selection process, whilst traditional project selection may be conducted individually. Further to the research done by Zhu et al. (2015) and Liu et al. (2017), a new acquisition method of belief distributions will be proposed in this paper, as the fact of employing a consistent selection process and having a large number of proposals makes it possible to collect experimental data for generating likelihoods. As the evaluation of R&D projects is often carried out by experts with subjective judgments, experimental data can be used to acquire evidence to represent assessments of experts. In this way, both real experimental data and experts' assessments can be used to support R&D project selection. The ER rule employs a belief structure to model various types of uncertainty and an inference rule for aggregating information, where conflicting pieces of evidence can be combined. As such, this paper utilizes combination of the assessments by experts to evaluate projects objectively and correctly.

## 3. The proposed model

### 3.1 A data-driven inference model

A data-driven inference model for multi-criteria R&D project selection mainly includes evidence acquirement and inference. The relationship between likelihood and basic probability (or degree of belief) as established in the ER rule provides a rigorous yet practical way to acquire evidence from experimental data (Yang and Xu, 2014).

Then information aggregation can be achieved by implementing the ER rule. The aggregation module consists of two parts, namely multi-criteria aggregation and multi-experts aggregation. The multi-criteria aggregation is associated with the relative weight of each criterion. The multi-experts aggregation is associated with both weight and reliability of each expert. Reliability is the inherent property of the evidence and refers to the quality of the information source, and nevertheless weight can be subjective and can be determined according to the individual who uses the evidence (Chen et al., 2015). The data-driven inference model is shown in Fig. 1.

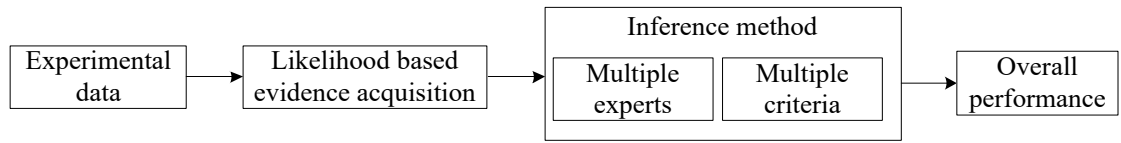

Fig. 1. A data- driven inference model

The model proposed in this paper comprises the following steps. First of all, a R&D selection problem is modelled using a belief structure, in which degrees of belief are acquired from historical data to represent experts' assessments. Experts' assessments can be the evaluation grade on each evaluation criterion, such as "poor, average, good and excellent". The ER rule is then employed to combine the assessments of multiple experts on multiple criteria, so as to generate an aggregated expert assessment for a project proposal. Finally, the aggregated expert assessment is used to provide a ranking for all R&D project proposals, to support the selection of the most favourable R&D proposals. The main components of the proposed model are described with details in the following sections.

### 3.2 The ER rule for inference

Suppose $\Theta = \{\theta_1, \ldots, \theta_N\}$ is a set of mutually exclusive and collectively exhaustive hypotheses or propositions, which is called the frame of discernment. The power set of $\Theta$, denoted by $2^\Theta$ or $P(\Theta)$, consists of $2^N$ subsets of $\Theta$ as follows

$$2^\Theta = \{\phi, \theta_1, \ldots, \theta_N, \{\theta_1, \theta_2\}, \ldots, \{\theta_1, \theta_N\}, \ldots, \Theta\}$$

A basic probability assignment (bpa) is a function $p_i: 2^\Theta \to [0, 1]$, satisfying

$$p_i(\phi) = 0$$

$$\sum_{\theta \subseteq \Theta} p_i(\theta) = 1$$

where $\phi$ is an empty set, $\theta$ is any subset of $\Theta$. $p_i(\theta)$ is a basic probability (or degree of belief) that is assigned exactly to $\theta$ but not to any of the subset of $\theta$.

In the ER rule framework, a piece of evidence $e_i$ is profiled by a belief distribution as follows

$$e_i = \left\{ (\theta, p_{\theta,i}), \forall \theta \subseteq \Theta, \sum_{\theta \subseteq \Theta} p_{\theta,i} = 1 \right\}$$

where $p_{\theta,i} = p_i(\theta)$ denotes the basic probability that evidence $e_i$ points exactly to proposition $\theta$, which can be any element of the power set $P(\Theta)$ except for the empty set. $(\theta, p_{\theta,i})$ is an element of evidence $e_i$, and it is referred as a focal element of $e_i$ if $p_{\theta,i} > 0$.

Suppose $w_i$ and $r_i$ are the weight and reliability of evidence $e_i$ respectively. Both of weight and reliability are in the range of $[0, 1]$. If two pieces of evidence $e_1$ and $e_2$ are independent of each other and defined by Equation (4) with weight ($w_1$ and $w_2$) and reliability ($r_1$ and $r_2$), the combined degree of belief to which $e_1$ and $e_2$ jointly support proposition $\theta$, denoted by $p_{\theta,e(2)}$, can be generated as follows

$$p_{\theta,e(2)} = \begin{cases} 0, & \theta = \phi \\ \dfrac{\widehat{m}_{\theta,e(2)}}{\sum_{D \subseteq \Theta} \widehat{m}_{D,e(2)}}, & \theta \neq \phi \end{cases}$$

$$\widehat{m}_{\theta,e(2)} = [(1-r_2)m_{\theta,1} + (1-r_1)m_{\theta,2}] + \sum_{B \cap C = \theta} m_{B,1} m_{C,2}$$

where $m_{\theta,i} = w_i p_{\theta,i}$ for any $\theta \subseteq \Theta$. There are two parts in the combination equation of the ER rule. The first square bracket part is the bounded sum of individual support for proposition $\theta$ from each of the two pieces of evidence, and the second part is the orthogonal sum of collective support for $\theta$ from both pieces of evidence.

The ER rule is the further development of D-S theory and the ER approach. D-S theory mainly has two rules. One is Dempster Rule and the other is Shafer's discounting rule. Dempster rule assumes the above reliability $r_i = 1$, while in the ER rule, $r_i$ can be any value between 0 and 1. It means that Dempster rule is a special case of the ER rule when $r_i = 1$. Shafer's discounting rule assumes $r_i$ can be any value between 0 and 1, but it assigns residual belief $(1 - r_i)$ to the frame of discernment (or global ignorance), while the ER rule assigns the residual belief $(1 - r_i)$ to the power set for redistribution after all pieces of evidence are combined. By assigning residual belief $(1 - r_i)$ to the frame of discernment, it disqualifies Shafer' rule as a probabilistic rule, while both Dempster rule and the ER rule are probabilistic rules.

The difference between the ER rule and the ER approach for multiple criteria decision analysis is that in the ER rule, weight $w_i$ and reliability $r_i$ are normally not equal and not normalised, while in the ER approach, $w_i = r_i$ and $w_i$ is normalised. Examples and the properties of the ER rule are discussed in details in Yang and Xu (2013).

### 3.3 Reliabilities

Each piece of evidence can be associated with a reliability, which refers to the quality of the information source to provide correct solution for a given problem (Smarandache et al., 2010; Yang and Xu, 2014; Liao et al., 2018). In the context of R&D project selection, the reliability of each piece of evidence provided by an expert can be measured to some extent by his past review performance if the expert has reviewed a number of projects in the past. Overall accuracy is not an appropriate metric to represent the real performance of an expert, because it can yield misleading results if the data set is unbalanced (that is, when the number of projects in "Funded" and "Unfunded" classes vary greatly). For example, if there were 16 unfunded projects and only 4 funded projects in the data set, an expert could easily be tempted to classify all samples as "Not fund". The overall accuracy for such classification would be 80%. In reality, the expert would have an impressive 100% recognition rate for the "Unfunded" category but a terrible 0% recognition rate for the "Funded" category. To make more detailed analysis than mere proportion of overall accuracy, a confusion matrix (Provost and Kohavi, 1998) is proposed to measure the reliability of evidence, as shown in Table 1.

Table 1. A confusion matrix for measuring the reliability of evidence

|  |  | Expert's recommendations | |
|---|---|---|---|
|  |  | Fund | Not fund |
| Actual outcomes | Funded | True Positive (*TP*) | False Negative (*FN*) |
|  | Unfunded | False Positive (*FP*) | True Negative (*TN*) |

In the above confusion matrix, of the "Funded" projects, an expert recommended that *TP* projects should be funded and *FN* should not, and of the "Unfunded" projects, he recommended that *FP* projects should be funded and *TN* should not. The true positive rate and the true negative rate can be formulated by

$$\text{True positive rate} = \frac{TP}{TP + FP}$$

$$\text{True negative rate} = \frac{TN}{TN + FN}$$

The reliability of evidence provided by an expert can be generated by the true positive rate if he makes a "Fund" recommendation, and the true negative rate otherwise.

### 3.4 Likelihood based evidence acquisition

In this section, the method for constructing belief distribution for ER rule will be introduced, while more details can be found in the reference (Yang and Xu, 2014).

Suppose $e_0$ is a piece of old evidence representing a prior distribution in the frame of discernment $\Theta = \{h_1, \ldots h_N\}$, or

$$e_0 = \left\{(h_i, p_{i0}), i = 1, \ldots, N, \sum_{i=1}^{N} p_{i0} = 1\right\}$$

where $p_{i0}$ is the prior probability that is assigned to hypothesis $h_i$ in advance based on prior information $I_0$, or $p_{i0} = p(h_i|I_0)$.

Suppose $c_{ij} = p(e_j|h_i, I_0)$ with $\sum_{j=1}^{L} c_{ij} = 1$ for $i = 1, \ldots, N$ is the likelihood to which the $j^{th}$ test result $e_j$ is expected to occur, given that the $i^{th}$ hypothesis is true and prior information $I_0$ is available. With a new test result $e_1$, Bayesian inference can be used to generate posterior probability as follows

$$p(h_i|e_1, I_0) = \frac{p(e_1|h_i, I_0)p(h_i|I_0)}{\sum_{i=1}^{N} p(e_1|h_i, I_0)p(h_i|I_0)}$$

In the above equation, the combination of old evidence profiled as a probability distribution over the set of hypotheses and new evidence profiled as likelihoods over the set of test results for a given hypothesis is not symmetrical (Shafer, 1976; Yang and Xu, 2014). This asymmetry underpins Bayesian inference. It may be difficult to classify multiple pieces of evidence as old and new evidence, and it is desirable to represent both old and new evidence in the same format, so that they can be combined in any order.

As mentioned, in Bayesian inference, evidence is represented by prior probability distribution and likelihood function. In the ER rule, new evidence can be acquired from likelihood and also profiled as probability distribution, or more generally belief distribution. Suppose $p_{ij}$ is the degree of belief to which the $j^{th}$ test result points to hypothesis $h_i$. The test result can then be profiled as evidence $e_j$ as follows

$$e_j = \left\{(h_i, p_{ij}), i = 1, \ldots, N, \sum_{i=1}^{N} p_{ij} = 1\right\} \quad j = 1, \ldots, L$$

If all tests to generate likelihoods are conducted independently, the degree of belief $p_{ij}$ can be generated from likelihood $c_{ij}$ as follows (Yang and Xu, 2014)

$$p_{ij} = c_{ij} / \sum_{n=1}^{N} c_{nj}$$

The ER rule reduces to Bayes' rule if the degrees of belief are given by Equation (12), each piece of evidence is fully reliable, and basic probability is assigned to singleton hypothesis only.

From the above analysis, we can utilise experimental data to represent experts' assessment grades by using belief distributions in the set of final funding outcomes. It can reveal the mapping relationships between the evaluation grades and the final funding outcomes, and it can provide a more reliable and intuitive way to represent evidence. The above process can be summarized as follows:

Step 1: Collect and process all the experimental data or sample information of peer experts;

Step 2: Construct likelihood function with the collected data;

Step 3: Calculate belief matrix of peer experts by Equation (12);

Step 4: Extract the belief distributions in accordance with the initial evaluation grades by the expert.

## 4. A case study

### 4.1 Problem description and data

China, as a rising scientific research power (Liu et al., 2015a), has attracted global attention (Liu et al., 2015b). The National Science Foundation of China (NSFC) is the largest funding agency for supporting fundamental research in China, and one of its major tasks is to evaluate and select R&D projects (proposals) with great potential of scientific breakthrough or social impacts (Tian et al., 2005). In this section, we investigate the R&D project selection problem in the NSFC. The selection process in the NSFC is mainly carried out through seven steps as shown in Fig. 2 (Feng et al., 2011; Silva et al., 2014; Zhu et al., 2015). In general, the NSFC announces a call for proposals (CFP) first, together with the application guidelines. Proposals are then completed by principal investigators (PIs) and submitted by supporting organizations, such as higher educational organizations and research institutes. Submitted proposals can be collected, validated electronically, and then assigned to external experts for peer review. The external experts review the proposals based on their levels of expertise and professional experience, as well as in accordance with the rules of the funding agency. Then an aggregation scheme is employed to combine assessments from peer review and to rank and identify the proposals that ought to be funded. A final decision is made based on the ranking and judgment of the panel of experts.

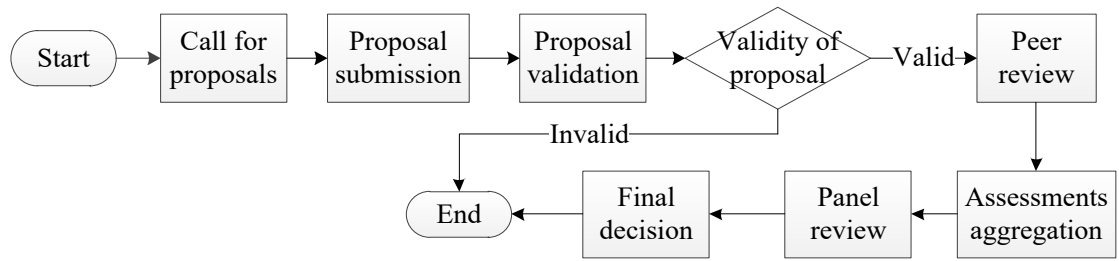

Fig. 2. The typical R&D project selection process in the NSFC

In the peer review step, division managers invite and assign external experts to evaluate projects on two criteria, namely "Comprehensive evaluation level" and "Funding recommendation". Four evaluation grades, including poor, average, good and excellent are used to assess the first criterion, while three grades, namely not fund, fund, and fund with priority, are used to assess the second criterion. The basic two criteria, denoted as $C_1$ and $C_2$ respectively, are used by experts to assess projects and $C$ is used to denote the overall performance of a project. Generally three to five experts working in the same or relevant fields are selected for assessing each project, denoted by $E_k$ ($k=1,2,…,K$; $K=3$, 4, or 5). The analytical process of the R&D project evaluation is shown in Fig. 3.

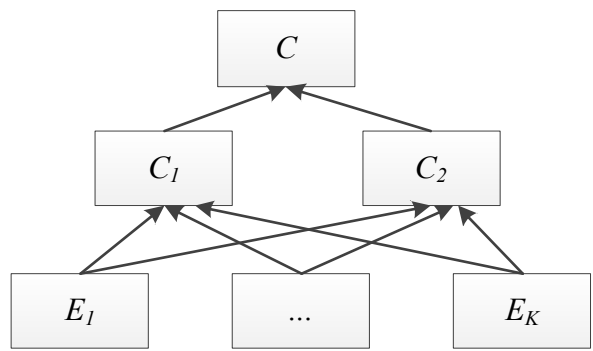

Fig. 3. The analytical process of the R&D project evaluation in the NSFC

To generate the overall performance of a proposal, an additive approach is currently employed to aggregate the assessments of proposals and rank them. Taking the management sciences department of the NSFC for example, two groups of values, which are 1, 2, 3, 4 and 0, 1, 2, are assigned to the evaluation grades of the two criteria respectively. The average scores of all experts on the two criteria are added together for initial ranking and providing support for panel review (Chen, 2009).

There are five funding instructions of programs for research promotion in NSFC:

general program, key program, major program, major research plan, and international (reginal) joint research program. The general program is one of the earliest funding instructions and is also widely known by researchers in China. In this case study, the review information and the project approval information of 497 projects were collected from the NSFC information center. All the 497 projects are from management sciences department on the general programs. As 78 projects are assessed by four experts and 419 projects are assessed by five experts, total 2407 assessments are generated as data samples. The approval outcome of a project, denoted by $\Theta$, can be divided into two categories, which are "Funded" ($\theta_1$) and "Unfunded" ($\theta_2$) respectively. The assessments and funding outcomes of the 497 projects on the two criteria are shown in Table 2 and Table 3.

Table 2. Experimental data for $C_1$

| Sample Data | Poor | Average | Good | Excellent | Total funding outcome |
|---|---|---|---|---|---|
| $\theta_1$ | 6 | 51 | 167 | 194 | 418 |
| $\theta_2$ | 260 | 900 | 629 | 200 | 1989 |
| Total evaluation | 266 | 951 | 796 | 394 | 2407 |

Table 3. Experimental data for $C_2$

| Sample Data | Not fund | Fund | Fund with priority | Total funding outcome |
|---|---|---|---|---|
| $\theta_1$ | 66 | 211 | 141 | 418 |
| $\theta_2$ | 1192 | 680 | 117 | 1989 |
| Total evaluation | 1258 | 891 | 258 | 2407 |

The likelihoods $c_{ij}$ can be generated from the above two tables using equations in Section 3.4 and the results are given by rows 3 and 4 of Table 4 and Table 5. As all the assessments to generate likelihoods are conducted independently, the degree of belief $p_{ij}$ can be generated from likelihood $c_{ij}$, as shown in rows 4 and 5 of Table 4 and Table 5.

Table 4. The causal belief matrix for $C_1$

| | | Poor | Average | Good | Excellent |
|---|---|---|---|---|---|
| $c_{ij}$ | $\theta_1$ | 0.0144 | 0.1220 | 0.3995 | 0.4641 |
| | $\theta_2$ | 0.1307 | 0.4525 | 0.3162 | 0.1006 |
| $p_{ij}$ | $\theta_1$ | 0.0989 | 0.2124 | 0.5582 | 0.8219 |
| | $\theta_2$ | 0.9011 | 0.7876 | 0.4418 | 0.1781 |

Table 5. The causal belief matrix for $C_2$

| | | Not fund | Fund | Fund with priority |
|---|---|---|---|---|
| $c_{ij}$ | $\theta_1$ | 0.1579 | 0.5048 | 0.3373 |
| | $\theta_2$ | 0.5993 | 0.3419 | 0.0588 |
| $p_{ij}$ | $\theta_1$ | 0.2085 | 0.5962 | 0.8515 |
| | $\theta_2$ | 0.7915 | 0.4038 | 0.1485 |

In Table 4, the value 0.0144 in the second row and third column represents the likelihood to which the assessment grade "Poor" is expect to occur given that the approval outcome of project is "Funded". The value 0.0989 in the fourth row and third column is the basic probability to which the approval outcome $\theta$ is believed to be $\theta_1$ if the project's "comprehensive evaluation level" is assessed to be "Poor".

**4.2 Illustration of the data-driven evidential reasoning rule based method**

The degree of belief that an assessment result points to a hypothesis is generated by implementing the above calculation. We use a particular project to illustrate how the proposed method can be applied for project evaluation. The project is evaluated by five experts $E_k$ ($k$=1,2,3,4,5) as shown in Table 6. The experts' reliabilities are generated by employing the proposed method in Section 3.3 with historical data for each of the five experts separately and their weights are assumed to be equal to their reliabilities. To facilitate the application of the ER rule, a MATLAB programme is developed for carrying out the numerical study described in this section.

Table 6. Original assessments by experts

| Experts | $C_1$ | $C_2$ | Reliability |
|---|---|---|---|
| $E_1$ | Excellent | Fund with priority | 0.6667 |
| $E_2$ | Good | Fund | 0.3466 |
| $E_3$ | Average | Not fund | 1.0000 |
| $E_4$ | Good | Fund | 0.2500 |
| $E_5$ | Good | Fund | 0.1000 |

The subjective assessments from experts can be represented by belief distributions. Expert $E_1$ assessed the project to be "Excellent" on the first criterion. From Table 4, we know that "Excellent" points to hypothesis $\theta_1$ with degree of belief of 0.8219 and points to the hypothesis $\theta_2$ with degree of belief of 0.1781. Then the subjective assessments on $C_1$ by $E_1$ can be represented by the following belief distributions: $\{(\theta_1, 0.8219), (\theta_2, 0.1781)\}$. The assessments for the project from the five experts can be transformed to probability distributions as shown in Table 7.

Table 7. Probability distributions by experts

| | $E_1$ | | $E_2$ | | $E_3$ | | $E_4$ | | $E_5$ | |
|---|---|---|---|---|---|---|---|---|---|---|
| | $C_1$ | $C_2$ | $C_1$ | $C_2$ | $C_1$ | $C_2$ | $C_1$ | $C_2$ | $C_1$ | $C_2$ |
| $\theta_1$ | 0.8219 | 0.8515 | 0.5582 | 0.5962 | 0.2124 | 0.2085 | 0.5582 | 0.5962 | 0.5582 | 0.5962 |
| $\theta_2$ | 0.1781 | 0.1485 | 0.4418 | 0.4038 | 0.7876 | 0.7915 | 0.4418 | 0.4038 | 0.4418 | 0.4038 |

With reliabilities given in Table 6, the joint support for the project by multiple experts can be generated by applying the ER rule. The weights of the assessments are supposed to be equal to their reliabilities. The five experts' assessments on each criterion are firstly aggregated and then the aggregated results on the two criteria are further aggregated to generate overall assessment on the top criterion. The aggregation results are shown in Table 8. It should be noted that the weight of the first criterion $w_1$ is assumed to be twice as much as the weight of the second criterion $w_2$ according to the regulation of NSFC.

Table 8. Aggregation results

|  | $C_1$ | $C_2$ | $C$ |
|---|---|---|---|
| $\theta_1$ | 0.3661 | 0.3909 | 0.3535 |
| $\theta_2$ | 0.6339 | 0.6091 | 0.6465 |

The overall performance of the project based on experts' assessments is generated. The funding decision of the project can be made by comparing the funding probability of the project to other projects.

**4.3 Comparative analysis**

A group of 100 projects from data samples is used for comparative analysis between the proposed inference model and the existing method in the NSFC and check whether their results are consistent with the actual funding results. Among the 100 projects, there are 20 funded projects and 80 unfunded ones. The histogram of the number of funded and unfunded projects on the scores under the existing method are visualised in Fig. 4.

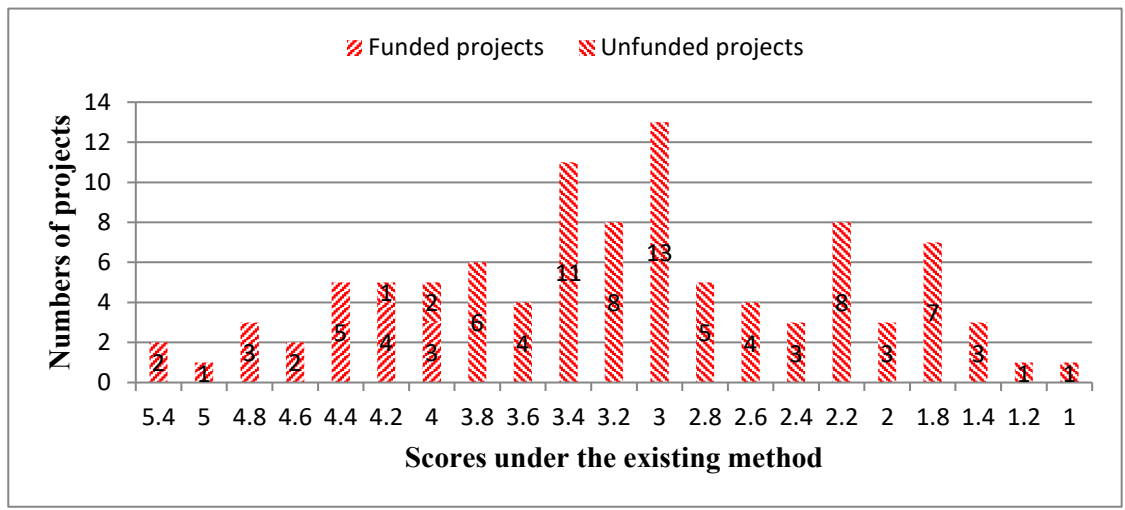

Fig. 4. The histogram distributions under the existing method

It can be seen from Fig.4 that the scores range of the 100 projects is [1.0, 5.4] and only 21 scores are used to map the performance of those projects. The limited number of scores lacks the ability to distinguish all the proposals. It can also be seen from Fig.4 that having the same score of 4.2, four projects were funded, but the other one was not. It is a similar situation with the score of 4. It should be noted that the figure looks great as the actual funding decision was largely based on the additive approach.

In Table 9, the proposed method in this paper is compared with the additive approach used by NSFC. As there were 20 funded projects in the data, the actual outcomes of top 20 projects are used to evaluate the inference accuracy.

Table 9. Actual outcomes for top 20 projects under the two methods

|  | Funded | Unfunded | Undifferentiated | Total number |
|---|---|---|---|---|
| The existing method | 17 | 1 | 2 | 20 |
| The proposed method | 18 | 2 | 0 | 20 |

The proposed model using the ER rule can provide superior inference performance, with higher funded projects and lower undifferentiated projects than the existing method. Under the existing method, 18 projects can be differentiated, including 17 funded projects and 1 unfunded project. The remaining 2 projects, which are from top 19 to top 20, have the same scores as the other 5 projects, as shown in Fig. 4. It means that the rest two projects have to be chosen from 5 projects with the same score of 4. The additive approach in the NSFC generates only twenty-six or fewer possible scores from five experts, and it lacks the ability to distinguish hundreds of projects.

The main difference between the existing method and the proposed method results from the ten projects, including the 5 projects with the score of 4.2 and the other 5 projects with the score of 4.0. The scores generated using the existing method and funding probabilities generated using the proposed method are shown in Table 10 headed by $x$ and $y$ respectively. The rankings of the projects are headed by $O_x$ and $O_y$ respectively.

Table 10. Results of the ten projects

| Project | $x$ | $O_x$ | $y$ | $O_y$ | Funded |
|---|---|---|---|---|---|
| 1 | 4.2 | 6 | 0.7002 | 1 | yes |
| 2 | 4.2 | 6 | 0.5821 | 3 | yes |
| 3 | 4.2 | 6 | 0.2973 | 6 | yes |
| 4 | 4.2 | 6 | 0.2012 | 10 | yes |
| 5 | 4.2 | 6 | 0.2864 | 7 | no |
| 6 | 4.0 | 7 | 0.6019 | 2 | yes |
| 7 | 4.0 | 7 | 0.3510 | 4 | yes |
| 8 | 4.0 | 7 | 0.3005 | 5 | yes |
| 9 | 4.0 | 7 | 0.2691 | 8 | no |
| 10 | 4.0 | 7 | 0.2297 | 9 | no |

In Table 10, $y$ represents the funding probabilities according to reviewers' recommendations. The figures are low. According to the synthesis evidence report of the international evaluation of funding and management of the NSFC, the attendance rate is 130%~160%, which means that over 130 percentages of projects will be sent to the panel review step. The probability of getting funded for projects at the panel review step will be 62.5%~76.92% if we take into this step account only. The funding probabilities of projects generated by reviewers' assessments can be lager when aggregated with the probability at the panel review step.

It is difficult to differentiate Project 1 from Project 5 under the existing method as they have the same score of 4.2, and the same problem happens for projects from Project 6 to Project 10. The proposed method avoids such a problem and can differentiate projects effectively. It should also be noted that the rankings of some projects are changed. Project 5 ranks higher than Project 6, 7 and 8 under the existing method. The ranking is reversed under the proposed method and is consistent with the actual funding outcomes. To reveal the reason of different results under the two methods, the original assessments of Project 5 and Project 6 provided by the experts and the reliabilities are listed in Table 11.

Table 11. Review information of Project 5 and Project 6

|  |  | $E_1$ | $E_2$ | $E_3$ | $E_4$ | $E_5$ |
|---|---|---|---|---|---|---|
| Project 5 | $C_1$ | Good | Good | Average | Excellent | Excellent |
|  | $C_2$ | Fund | Fund | Not fund | Fund with priority | Fund |
|  | $R$ | 0.4286 | 0.3478 | 1.0000 | 0.3478 | 0.2857 |
| Project 6 | $C_1$ | Good | Good | Good | Average | Excellent |
|  | $C_2$ | Fund | Fund | Fund | Fund | Fund |
|  | $R$ | 0.3478 | 0.4000 | 0.6667 | 0.3478 | 0.2500 |

If only assessments are taken into account, Project 5 has better assessments than Project 6, as shown in the second, third, fifth and sixth rows of Table 11. Results have changed when taking reliabilities into account. As shown in the fourth row of Table 11, the reliability of $E_3$ for Project 5 is 1.0000. The expert is regarded as more reliable and his negative recommendation plays a more important role in the aggregation process. So Project 5 gets a lower score than Project 6 under the proposed method with expert reliabilities.

It's worth pointing out that one outcome of the proposed method is not consistent with the actual funding result, which is Project 4. The original assessments of Project 4 provided by the experts and the reliabilities are listed in Table 12.

Table 12. Review information of Project 4

|  |  | $E_1$ | $E_2$ | $E_3$ | $E_4$ | $E_5$ |
|---|---|---|---|---|---|---|
| Project 4 | $C_1$ | Excellent | Good | Good | Excellent | Average |
|  | $C_2$ | Fund with priority | Fund | Fund | Fund | Not fund |
|  | $R$ | 0 | 0.2981 | 0.2981 | 0 | 0.9834 |

As can be seen from Table 12, experts gave high recommendations for Project 4, especially $E_1$ and $E_4$. Only assessments of experts are taken into account in the existing method and Project 4 got a high score and was funded. The reliabilities of the two experts are both 0, the outcome value of the proposed method for Project 4 is low when taking reliabilities into account.

## 5. Discussion

### 5.1 Generalizations of the newly proposed model

In the preceding section, the newly proposed model has been demonstrated for the NSFC project selection. To guide managers for better application and gain wider acceptance for practice, it is necessary to generalize the proposed model with a simple example.

The model follows these steps:

(1): Collection of experts' assessments. For each project selection problem, a certain set of evaluation grades would be suitable to capture the performance of candidate projects. Suppose H = $\{\theta_1, ..., \theta_5\}$ = {Worst, Poor, Average, Good, Best}, two independent experts are invited to make recommendations for a given (2): project and choose 'Poor' and 'Good' respectively.

(2): Representation of experts' assessments with belief distributions. Refer to the likelihood based evidence acquisition process mentioned in Section 3.4, managers can extract the corresponding belief distributions. Suppose the two assessments are represented by the following belief distributions:

'Poor': $\{(\theta_1, 0.2), (\theta_2, 0.8)\}$

'Good': $\{(\theta_1, 0.7), (\theta_2, 0.3)\}$

(3): Aggregation of experts' assessments. Before aggregation, managers should collect historic evaluation performance data of experts to calculate their reliabilities. With reliabilities and weights of experts, the aggregated funding probabilities can be calculated by using the ER rule. Suppose reliabilities are equal to weights for both two experts, and values are 0.25 and 0.85 respectively. Then the aggregation results of the two assessments should be $\{(\theta_1, 0.633), (\theta_2, 0.367)\}$.

(4): Aggregation of information for different attributes. If the selection process includes several attributes, this step will be needed. Managers should determine relative importance of all the attributes and aggregate information for generating overall funding probabilities.

(5): Decision making by comparison and ranking. By ranking and comparing the funding probabilities of all candidate projects, managers can choose which one or ones to further implement depend on practical needs and budgets.

The order of step 'Aggregation of experts' assessments' and step 'Aggregation of information for different attributes' is interchangeable according to practical needs. All these steps can be computerized and tooled to make the whole process user-friendly.

The proposed model in this paper can also be applied to non-R&D project selection in any industries, governmental organizations or companies, if they are featured with similar evaluation structure, criteria and process.

**5.2 Benefits of the newly proposed model**

There are a number of benefits in the application of the ER rule based inference model to the R&D selection process.

One of the benefits is the interpretability and fairness of the selection results. Due to differences in their ability to provide correct assessments, experts play different roles in the newly proposed model. It completely depends on expert's judgments and assessments, and an interpretable and fair selection process gives further guidance for proposals which fail to receive a funding. Applicants who fail to receive a funding can differentiate assessments by various experts, and focus on those suggestions that are provided by more reliable experts to improve their proposals further. According to the NSFC Regulations, peer experts shall be brought into full play. Funding decisions for R&D projects should be based on peer experts' judgements other than subjective and politically biased influence. Even those subjective and politically loaded influences exist, the proposed model takes experts' reliabilities into account and those influences will be reduced substantially.

Another benefit is the efficiency improvement of the selection process. Due to the rapid rising of emerging countries such as China, many R&D projects conducted nationwide increase the workload of the selection process for funding agencies including the NSFC. The existing method requires excessive time expenditure by experts, as it has many projects of same score and those undifferentiated projects need to be analyzed further for panel review. The performances of projects are distinguished well in comparison with the existing method. The time-consuming panel review work for both managers and experts will be reduced to a minimum, and the quality and efficiency can be improved significantly.

A multiple criteria decision analysis software tool can be designed and used to conduct project selection process. With appropriate technical support, it can reduce time cost for managers and further improve efficiency of project selection.

## 6. Conclusions

In this paper, we proposed a data-driven inference model to analyze multiple criteria R&D projects selection problems. The relationship between likelihoods and basic probabilities was explored to acquire evidence from the subjective assessments of experts as recorded in historical data. The ER rule based inference model was constructed to generate the combined belief degrees for distributed assessment and also the funding probabilities for ranking of alternatives. It utilized experimental data to represent experts' assessments by using belief distributions over the set of final funding outcomes, and through this historic statistics it helps experts and applicants to understand the funding probability of a given assessment grade. It can provide a way to make fair decisions by taking experts' reliabilities into account, since it encourages experts to keep objective and provide fair evaluation for improving their reliabilities and reputation. To illustrate the contribution of this research in a practical sense, a R&D project evaluation and selection framework was incorporated with the data-driven inference model for the implementation to a real-world case study. The effectiveness of the proposed model was validated by comparing the outcomes of the proposed model with the outcomes of the existing methods in the NSFC.

The proposed multi-criteria and multi-experts decision making model can be a useful tool for funding agencies to tackle R&D project evaluation and selection problems when experimental data are available. In addition, the reliabilities of experts in the proposed model can be seen as a criterion to estimate the qualification of an expert and can be used to assign appropriate experts for evaluating a project. It should be noted that the calculation of the mapping relationships between the evaluation grades and the final funding outcomes under study in this paper is based on the historical data of all the experts instead of calculating them for each expert, since an appropriate set of historical data are required to obtain the reasonable outcomes. Moreover, the application to other funding agencies or industries can be conducted to show the generalization of the proposed model.

**Acknowledgements**

This research is partially supported by Zhejiang Provincial Natural Science Foundation of China (Grant numbers LQ18G010005 and LQ18G030010) and the National Natural Science Foundation of China (Grant numbers 71801189, 71373065, 91746202 and 71433006).